\documentclass[aps,prd,twocolumn,showpacs,preprintnumbers,amsmath,amssymb,floatfix,10pt]{revtex4}
\usepackage{graphicx}
\usepackage{subfig}
\usepackage{dcolumn}
\usepackage{float}
\usepackage{bm}
\usepackage{enumerate}
\usepackage[english]{babel}
\usepackage{amsfonts}
\usepackage{amsthm}
\usepackage{amssymb}
\usepackage{calc}
\usepackage{ifthen}
\usepackage{color}

\usepackage{inputenc}
\inputencoding{utf8}

\newcommand{\numset}[1]{\mathbb{#1}}

\newcommand{\integers}{\numset{Z}}
\newcommand{\reals}{\numset{R}}
\newcommand{\complexes}{\numset{C}}
\renewcommand{\Re}{\text{Re}}
\renewcommand{\Im}{\text{Im}}
\newcommand{\sphere}{\mathsf{S}}
\newcommand{\abs}[1]{\left\lvert#1\right\rvert}

\newcommand{\norm}[1]{\lVert#1\rVert}

\newcommand{\doo}{\partial}
\newcommand{\doomu}{\doo_\mu}
\newcommand{\dooMU}{\doo^\mu}
\newcommand{\density}[1]{\mathcal{#1}}
\newcommand{\half}{\tfrac{1}{2}}
\DeclareMathOperator{\grad}{\nabla}
\newcommand{\crossprod}{\times}
\newcommand{\dotprod}{\cdot}
\newcommand{\diff}{\text{d}}
\renewcommand{\exp}[1]{e^{#1}}

\newcommand{\R}{{\mathbb{R}}}

\newcommand{\beq}{\begin{equation}}
\newcommand{\eeq}{\end{equation}}
\newcommand{\bea}{\begin{eqnarray}}
\newcommand{\eea}{\end{eqnarray}}
\newcommand{\ra}{\rightarrow}

\newcommand{\cd}{\partial}

\newcommand{\ignore}[1]{}

\newcommand{\xv}{\mbox{\boldmath{$x$}}}
\newcommand{\yv}{\mbox{\boldmath{$y$}}}
\newcommand{\zv}{\mbox{\boldmath{$z$}}}
\newcommand{\dv}{\mbox{\boldmath{$d$}}}
\newcommand{\nv}{\mbox{\boldmath{$n$}}}

\begin{document}

\title{Easy plane baby skyrmions}

\author{Juha J\"aykk\"a}
\email{juhaj@iki.fi}
\author{Martin Speight}
\email{j.m.speight@leeds.ac.uk}
\affiliation{
School of Mathematics,
University of Leeds,
LS2 9JT, Leeds, United Kingdom
}

\date{\today}

\begin{abstract}
  The baby Skyrme model is studied with a novel choice of potential, $V=\half \phi_3^2$. This ``easy plane''
  potential vanishes at the equator of the target two-sphere. Hence, in contrast to previously studied cases,
  the boundary value of the field breaks the residual $SO(2)$ internal symmetry of the model. Consequently,
  even the unit charge Skyrmion has only discrete symmetry and consists of a bound state of two half lumps. A
  model of long-range inter-Skyrmion forces is developed wherein a unit Skyrmion is pictured as a single
  scalar dipole inducing a massless scalar field tangential to the vacuum manifold. This model has the
  interesting feature that the two-Skyrmion interaction energy depends only on the average orientation of the
  dipoles relative to the line joining them.  Its qualitative predictions are confirmed by numerical
  simulations. Global energy minimizers of charges $B=1,\ldots,14,18,32$ are found numerically. Up to charge
  $B=6$, the minimizers have $2B$ half lumps positioned at the vertices of a regular $2B$-gon. For charges $B
  \ge 7$, rectangular or distorted rectangular arrays of $2B$ half lumps are preferred, as close to square as
  possible.
\end{abstract}

\maketitle

\section{\label{sec:intro}Introduction}

The Skyrme model, and its smaller relative, the baby Skyrme model, have received much attention during the
last few decades. The models are very interesting from the mathematical physics point of view: they contain
topologically nontrivial field configurations called topological solitons and the Skyrme model arises as a
large-$N$ limit of QCD \cite{Witten:1983tw,Witten:1983tx}. The baby Skyrme model is the two-dimensional analog
of the full model and its main interests lie in its easier tractability, the possibility of using the baby
Skyrme model results as guidelines in investigating the full Skyrme model and some condensed matter contexts,
where the baby Skyrme model itself is of physical relevance \cite{1993PhRvB..4716419S}. The baby Skyrme model
was investigated intensively during the 1990s but interest in the model has recently received a new boost
\cite{2008PhRvE..77c6612H, 2008PhRvD..77k6002H, 2009TMP...160..933H, 2008PhRvD..77e4009H, 2008Nonli..21..399H,
  2009PhRvD..80j5013A, 2010PhRvD..81h5007A, 2010JPhA...43N5201S}, apparently due to the prevalence of baby
Skyrmions in various condensed matter systems \cite{2006Natur.442..797R+2006cond.mat..3104R}, such as
Fe$_{1-x}$Co$_{x}$Si \cite{2010Natur.465..901Y}, MnSi \cite{2009Sci...323..915M, 2009PhRvL.102r6602N} and
quantum Hall systems (for a review, see \cite{2009RPPh...72h6502E} and references therein), despite the fact
that these are not modeled by the baby Skyrme model. It is against this background that we investigate the
baby Skyrme model.

The behavior of the solitons of the baby Skyrme model depends crucially on the choice of potential. For some
potentials, there will be no smooth solutions at all; for some, like that studied in
\cite{Nonlinearity_1990_3_773}, there will be exact solutions for the equations of motion and for the rest,
such as here, numerical methods must be used. So far, only cases where the potential and boundary condition
preserve the $SO(2)$ symmetry of the Lagrangian have been studied in detail. Here we study the easy plane
potential, which breaks $O(3)$ symmetry to $SO(2)$, and the boundary condition further breaks the symmetry to
$\integers_2$. This endows the model with one massive and one massless mode and one expects significantly
different behavior from $SO(2)$ symmetric cases.

We use numerical methods to demonstrate that the model has a rich static energy structure, including local
minima and that multi-Skyrmions are stable in this model. We also develop an approximate way to estimate the
long-range interaction of two unit baby Skyrmions and verify this numerically. We identify some open questions
and present a conjecture about the energy of the global minimum, when the charge approaches infinity.

\section{\label{the_model}The model}

The Lagrangian density of the model, defined on $\reals^{1,2}$, is
\begin{align}
  \label{eq:bs_model}
  \density{L} &= \half c_2\doomu \phi^a \dooMU \phi_a - \tfrac{1}{4} c_4 F_{\mu\nu} F^{\mu\nu} - c_0 \half U(\phi)^2,\\
  F_{\mu\nu} &= \epsilon_{abc} \phi^a \doomu \phi^b \partial_\nu \phi^c,
\end{align}
where the first terms are called the Dirichlet, the Skyrme and the potential terms, respectively and $c_i$ are
free parameters (coupling constants) of the model. The subscripts are chosen to match the number of
derivatives in the term and will prove useful later.

Choosing the usual metric $(+,-,-)$ yields the static energy density
\begin{align}
  \label{eq:bs_energy}
  \density{E} &= \half c_2 \norm{\grad \phi}^2 + \half c_4 (F_{12})^2 + \half c_0 U^2.
\end{align}
If one requires the solitons to have finite energy, it is necessary to impose the boundary condition
\begin{gather}
  \label{eq:usual_boundary}
  \lim_{r\to \infty}\phi(r) =: \phi_\infty = \text{ constant,} 
  \intertext{and choose $\phi_\infty$ from the vacuum manifold of the model, i.e. such that}
  U(\phi_\infty)=0.
\end{gather}

As usual, Eq.~\eqref{eq:usual_boundary} allows us to consider the field as map $\phi: \sphere^2 \to
\sphere^2$, which gives the system its interesting topological properties, namely, its fields are classified
topologically by the homotopy group $\pi_2(\sphere^2)$. The homotopy invariant of the field is called the
(topological) \emph{degree} and is given by the expression
\begin{gather}
  \label{eq:bogomolny}
  B = \frac{1}{4\pi} \int_{\reals^2} \phi \dotprod \left(\partial_1 \phi \crossprod \partial_2 \phi\right)
    \diff^2 x.
\end{gather}

Furthermore, one can expect the system to support topological solitons based on two factors. First, Derrick's
theorem of nonexistence does not apply since the terms zeroth and fourth order in derivatives behave in
opposite ways upon uniform scaling (the second order term is scale invariant), thus allowing solutions with a
preferred size. The second reason is the existence of a Bogomol'nyi bound for the (static) energy which
depends on the degree of the field, given by \cite{2010JPhA...43N5201S}
\begin{gather}
  E \ge 4\pi(c_2 + \sqrt{c_0 c_4}\langle U \rangle)\abs{B},
\end{gather}
where $\langle U \rangle$ is the average value of $U:\sphere^2 \to \reals$. Unlike in some models, the bound
cannot generically be saturated in this model.

In this work, we take $U(\phi) = \phi_3$ and $\phi_\infty = (1,0,0)$. This potential has the exceptional
effect of allowing the breaking of the symmetry of the energy almost completely. While the energy density
exhibits $SO(2)$ symmetry (from the potential), the choice of boundary condition breaks this down to
$\integers_2$. Hence, even the Skyrmion with unit topological charge will have only discrete symmetry.

We seek static solutions in various homotopy classes and investigate the forces between some of these. In
order to find a static solution, we numerically minimize the static energy of an initial field
configuration. Some initial configurations will be constructed from existing minima, but those that are
obtained from an exact expression are taken to be
\begin{gather}
  w = \lambda (x+iy)^B \\
  \phi = \frac{1}{\abs{w}^2+1}
  \begin{pmatrix}
    \abs{w}^2-1 \\
    2 \Re(w) \\
    -2 \Im(w)
  \end{pmatrix},
  \intertext{which for $B=1$ gives}
  \label{eq:initial}
  \phi = \frac{1}{\lambda^2(x^2+y^2)+1}
  \begin{pmatrix}
    \lambda^2(x^2+y^2)-1 \\
    2 \lambda x \\
    -2 \lambda y 
  \end{pmatrix},
\end{gather}
where $\lambda$ is simply a scale factor to adjust the size of the initial configuration. We also added random
ripples to destroy any discrete symmetries left over from the initial configuration. It was observed that this
is usually unnecessary, but some minimizers take different orientations (but keep their shape) when the random
ripples are omitted.

In all that follows, we set $c_0 = c_2 = c_4 = 1.0$.

\section{\label{sec:the_unit_skyrmion}The unit skyrmion}

The lattice approximation of derivatives is achieved by using a simple forward differencing scheme. The
discrete energy functional is then minimized using gradient based methods.

The gradients are computed from the discrete total energy in a straightforward manner. The energy minimization
is then achieved with the TAO\cite{tao-user-ref} and PETSc\cite{petsc-web-page,petsc-user-ref,petsc-efficient}
parallel numerical libraries. The libraries provide several different minimization schemes, of which we have
chosen the (in this case) fastest algorithm, the limited memory quasi-Newton algorithm (also called a variable
metric algorithm) with BFGS\cite{Broyden:1970aa,Fletcher:1970aa,Goldfarb:1970aa,Shanno:1970aa} formula for
Hessian approximations. We consider the minimization to have converged, when the $\sup$-norm of the gradient
is less than $10^{-7}$. This may sound like a very strong requirement, but it turns out to be necessary. Our
initial choice of $10^{-4}$ works for most cases, but when the results were checked with a the stronger
requirement, they experienced significant changes and decreases in energy after attaining the looser
condition. Using a still stronger criterion did not change the energies or shapes of the minimizers any
further.

The accuracy of the numerical scheme was tested against the known exact solution of $w=(x+iy)$
\cite{Nonlinearity_5_563-583} for the potential $U=(1-\phi_3)^2$. We find that the accuracy of the numerical
scheme is heavily dependent on the size of the lattice, due to the energy density of the Dirichlet term being
slowly decaying. When in a large enough lattice, the difference in energies between our minimizer and the
exact solution is less than $1\%$, so we expect similar accuracy in the present model as well, since it has
similar decay properties.

\begin{figure}[h]
  \centering
  \subfloat[][The energy density of the central region of the lattice.]{\includegraphics[width=4cm]{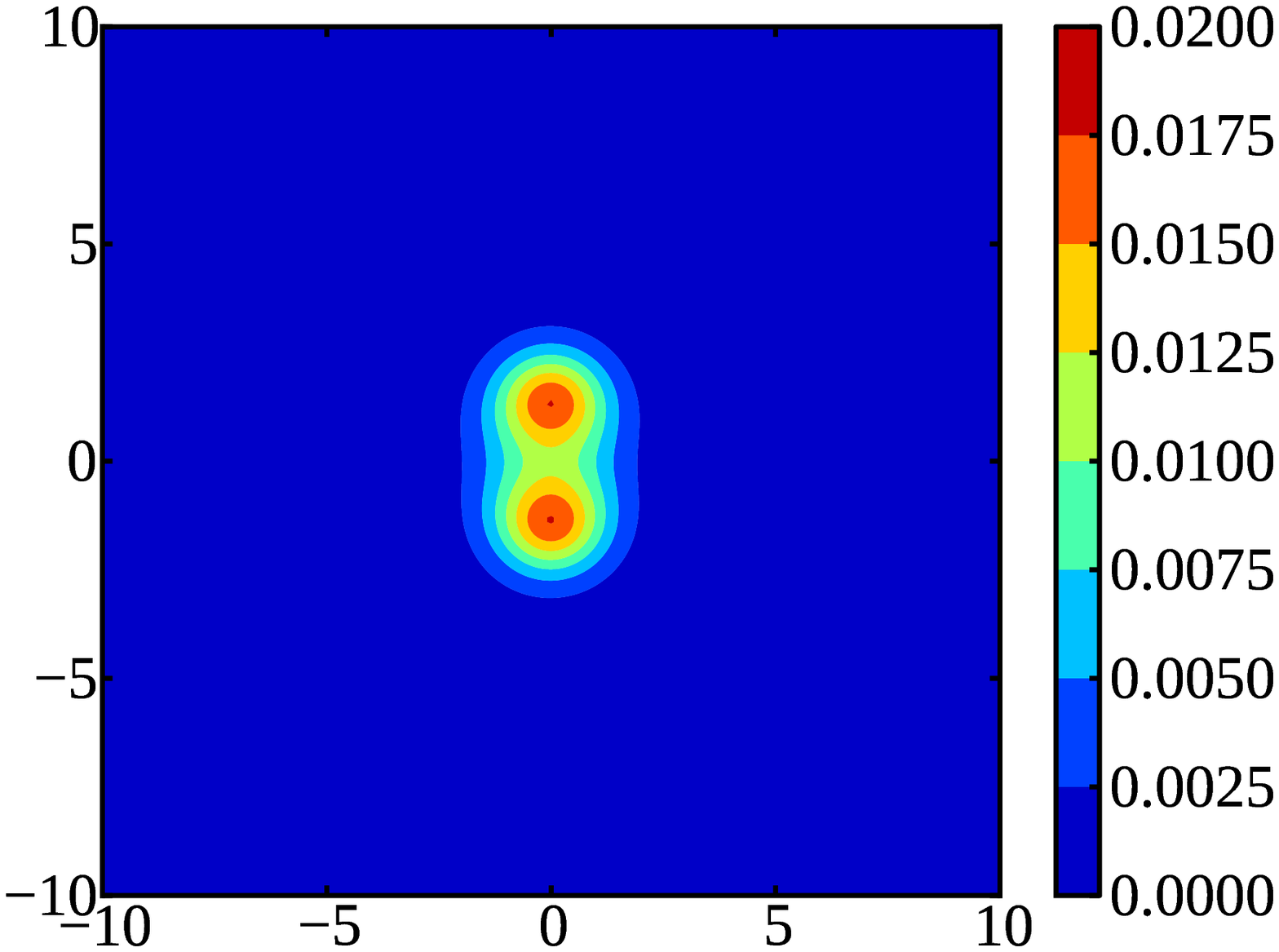}\label{fig:b1final}}
  \subfloat[][The contour plot of $\phi_3$.]{\includegraphics[width=4cm]{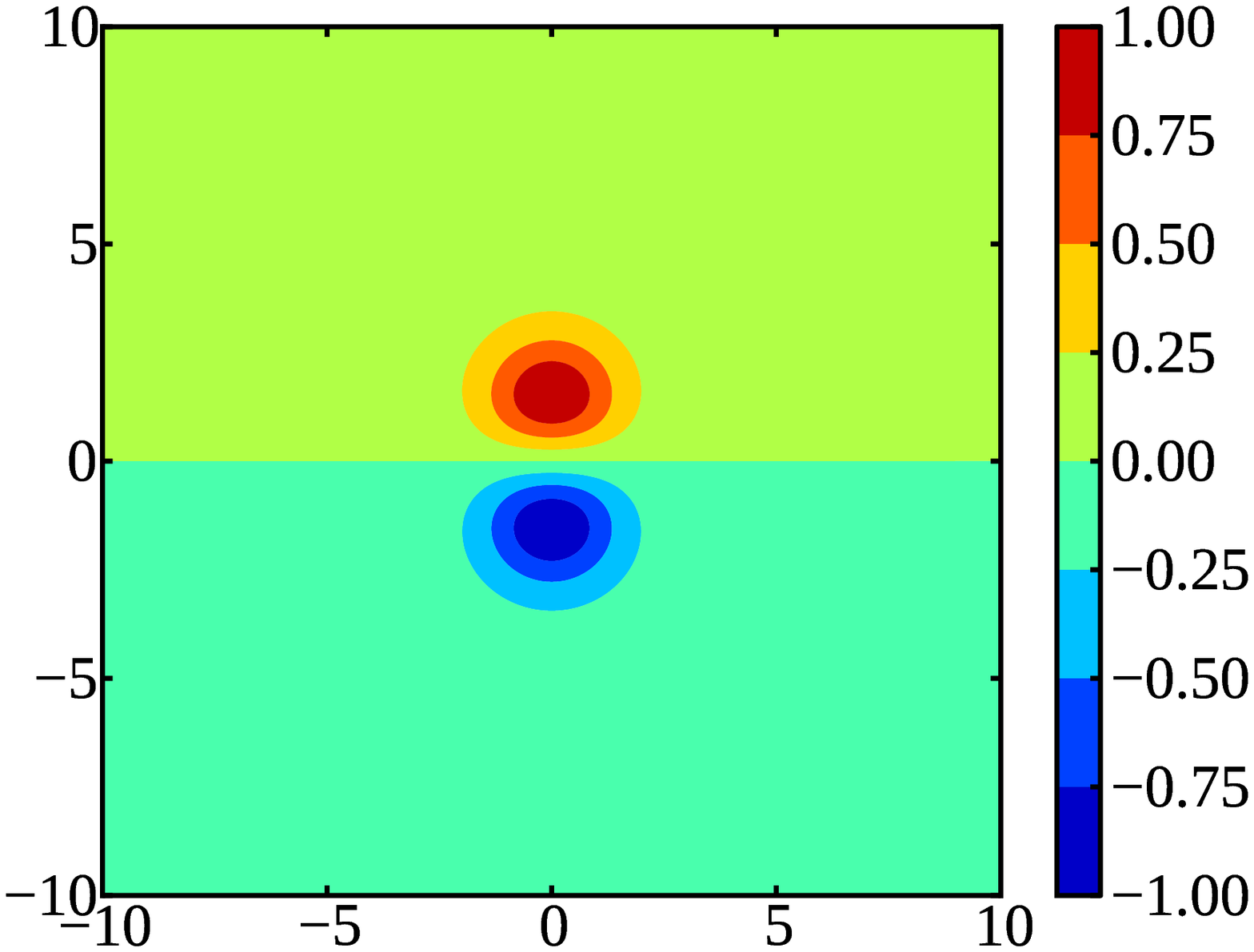}\label{fig:b1fields}}
  \caption{The minimizer for degree $B=1$.}
\end{figure}

We always use a square lattice and will refer to the number of lattice points along its side by a number $N$.
Starting with \eqref{eq:initial} on a $N=801$ lattice with lattice constant $h=0.1$, the corresponding minimum
energy density is depicted in Fig.~\ref{fig:b1final}. The broken rotational symmetry is immediately evident in
the form of the two red peaks. A view of how the field $\phi_3$ itself behaves is shown in \ref{fig:b1fields}
and an immediate observation is that the half-plane $y>0$ gets mapped to the upper hemisphere of the target
$\sphere^2$ and the half-plane $y<0$ gets mapped to the lower hemisphere. This, combined with the fact that
each individual peak in energy density strongly resembles a $\complexes P^1$ model lump, leads us to call the
solution on a half-plane a \emph{half lump}.

Note that the energy density image is zoomed in on the central area of the numerical lattice, where almost all
the energy density is concentrated. Obviously, the whole computation could have been done in a much smaller
lattice, but we opted not to find the optimal lattice sizes for each and every configuration due to the fact
that the amount of computer resources saved would have been small. We also confirmed that the resulting field
configuration is not dependent on either the value of the lattice constant or ``physical'' size of the
computational box by first repeating the computation with lattice constant $h=0.1$ and $N=1599$ and then
repeating again with $h=0.05$ and $N=1599$.

It is instructive to look at the fields themselves, in particular, their
asymptotics at large $r$. Since $\phi\ra(1,0,0)$ as $r\ra\infty$, we must have
\beq
\phi=(\sqrt{1-\phi_2^2-\phi_3^3},\phi_2,\phi_3)=(1,0,0)+(0,\phi_2,\phi_3)+\cdots
\eeq
at large $r$, where $\phi_2,\phi_3$ are small. Substituting this into ${\cal L}$
and keeping only leading order terms yields the Lagrangian density of an
uncoupled pair of real scalar fields,
\beq
{\cal L}=c_2\left\{\frac12\cd_\mu\phi_2\cd^\mu\phi_2
+\frac12\cd_\mu\phi_3\cd^\mu\phi_3-\frac12\frac{c_0}{c_2}\phi_3^2\right\}
\eeq
one massless ($\phi_2$) and the other of mass $\sqrt{c_0/c_2}$ ($\phi_3$).
One expects the large $r$ field of a $B=1$ soliton to be well approximated 
by a static solution of this linear model. Since $B=1$, $\phi_2+i\phi_3$ should
wind once around $0$ as one traverses a circle at large $r$. This leads one to
predict that
\beq
\phi_2(r,\theta)\sim\frac{q_2}{r}\cos\theta,\qquad
\phi_3(r,\theta)\sim q_3 K_1(\sqrt{c_0/c_2}r)\sin\theta
\eeq
at large $r$, where $q_2,q_3$ are unknown constants.

\begin{figure}[h]
  \centering
  \subfloat[][Plot of $r \phi_2(r,\theta)$ against $\theta$ and best fit for $q_2 \cos \theta$.]{\label{fig:b1masslessfit}\includegraphics[width=0.45\columnwidth]{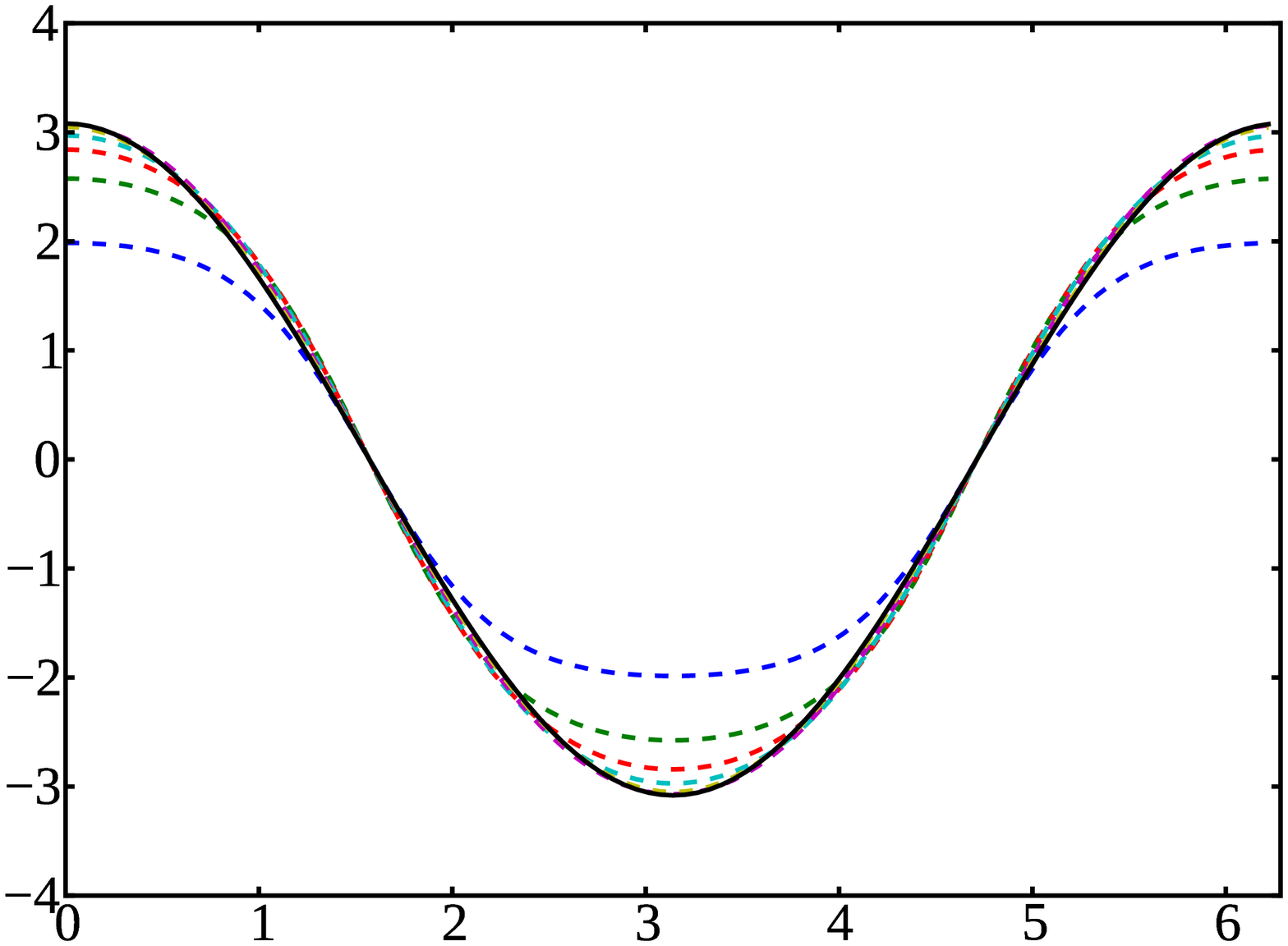}}
  \subfloat[][Plot of $\phi_3(r,\theta)/K_1(r)$ against $\theta$ and best fit for $q_3 \sin \theta$.]{\label{fig:b1massivefit}\includegraphics[width=0.45\columnwidth]{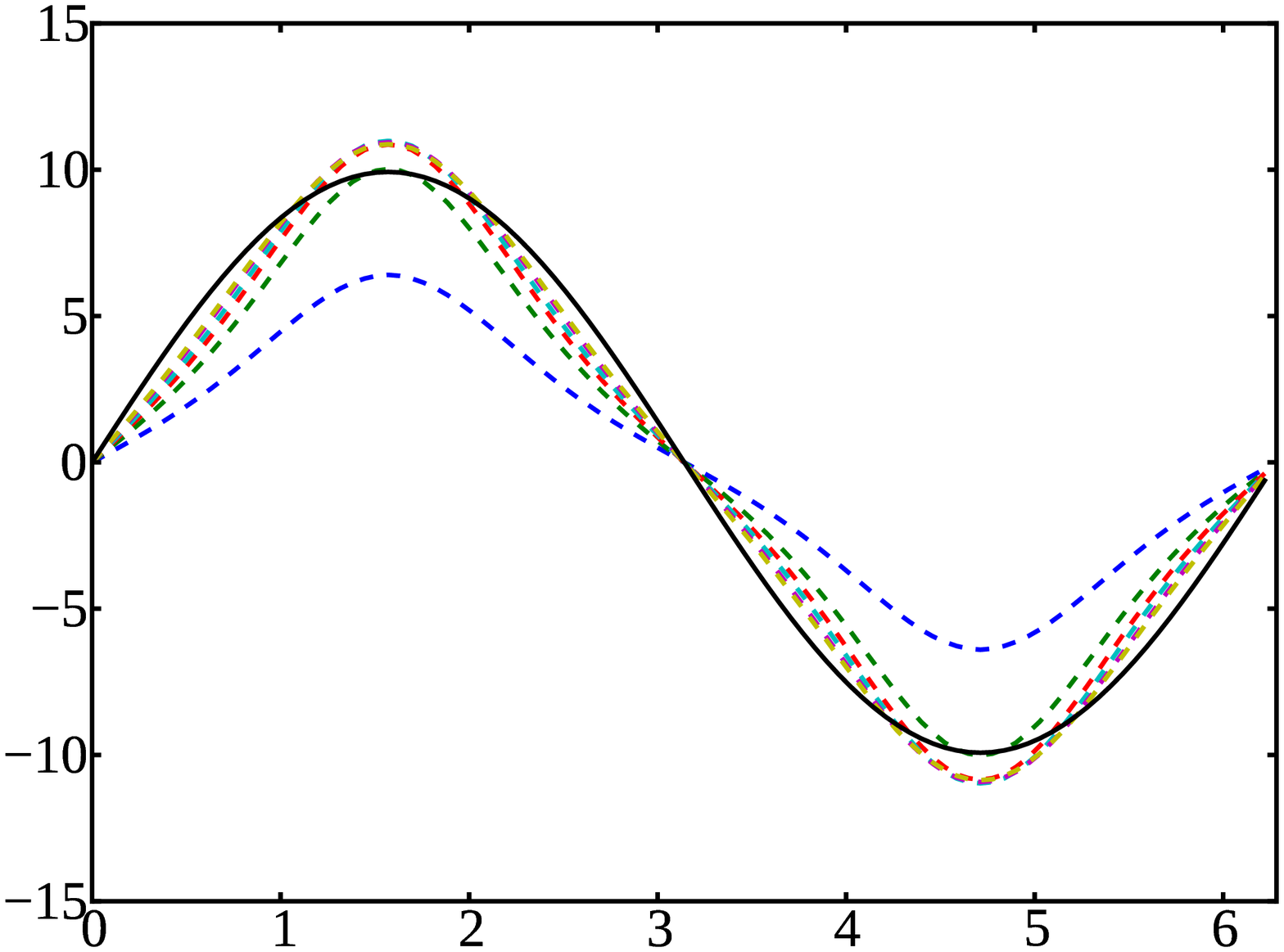}}
  \caption{Plots of $\phi_2$ and $\phi_3$ for $r \in \{2,3,4,5,7.5,10\}$ (the dashed lines, colored blue,
    green, red, cyan, magenta, and yellow, respectively) and the corresponding best fits (solid black).}
  \label{fig:b1fits}
\end{figure}

Numerical evidence in favor of the conjectured asymptotics for $\phi$ is presented in Fig.~\ref{fig:b1fits},
which shows plots of (\subref*{fig:b1masslessfit}) $r\phi_2(r,\theta)$ and (\subref*{fig:b1massivefit})
$\phi_3(r,\theta)/K_1(r)$ against $\theta$ for an increasing sequence of values of $r$. By fitting sinusoidal
curves, we find that $q_2 \approx 3.1$ and $q_3 \approx 9.9$. Note that, at large $r$, $K_1(r)$ decays like
$e^{-r}/\sqrt r$, so the massive field $\phi_3$ decays much more quickly than the massless field
$\phi_2$. This has important consequences for the long-range forces between Skyrmions, as we shall see in the
next section.

Having found the $B=1$ solution, it is natural to ask whether there are multisolitons, that is localized
solutions with $B>1$. Given the linear dependence of the energy bound on $B$, Eq.~\eqref{eq:bogomolny}, it is
conceivable that all these would consist of $B$ copies of the $B=1$ solution at infinite separation. Indeed,
this is what happens for the potential $U=(1-\phi_3)^2$, where the localized $B>1$ solutions are unstable
against decay to $B$ infinitely separated solutions with $B=1$\cite{Nonlinearity_1990_3_773}.

In that model (the so-called ``holomorphic'' model), the force between a pair of unit charge skyrmions turns
out to be repulsive, whatever their separation and relative orientation. By contrast, we will see that in the
current model (as for several other choices of potential) there is a choice of relative orientation for which
solitons attract one another at long range (the so-called attractive channel), making it plausible that
spatially localized solutions of every charge exist.

\section{\label{sec:inter_skyrmion_forces}Interskyrmion forces}

Intersoliton forces in the baby Skyrme model with an easy-axis potential
$\frac12U(\phi)^2=1-\phi_3$ were analyzed in detail by
Schroers et al in \cite{1995ZPhyC..65..165P}, and their methods readily adapt to the
current setting. We shall assume that the dominant interaction at
long range is mediated by the massless field $\phi_2$ and ignore the
contribution of the faster decaying field $\phi_3$. The key observation is that
the asymptotic field 
\beq
\phi_2(r,\theta)=\frac{q_2}{r}\cos\theta
\eeq
coincides with the static solution of the linearized model
\beq
{\cal L}=\frac12\cd_\mu\phi_2\cd^\mu\phi_2+\kappa\phi_2
\eeq
in the presence of an appropriate point source $\kappa$ at the origin, in
this case
\beq\label{sj}
\kappa(\xv)=-\frac{q_2}{2\pi}\cd_x\delta(\xv)
\eeq
where $\delta$ denotes the two-dimensional Dirac delta distribution. To
see this, note that the elementary kernel \cite{amp_part_1_1982} 
for Laplace's equation on
$\R^2$ is $G(\xv)=(2\pi)^{-1}\log|\xv|$, that is,
\beq
G_{xx}+G_{yy}=\delta(\xv).
\eeq
Source (\ref{sj}) induces
the field in standard orientation and position. The field rotated (spatially)
through angle $\alpha$ and translated to position $\yv\in\R^2$ is induced
by the source
\beq
\kappa(\xv)=\dv\cdot\nabla\delta(\xv-\yv)
\eeq
where $\dv=-q_2(\cos\alpha,\sin\alpha)$. Physically, one
interprets this as the massless
scalar field induced by a dipole source of moment $\dv$ at position
$\yv$. It is important to realize that the dipole moment which replicates
the soliton's asymptotic field points {\em orthogonally} to the line 
connecting
its constituent half lumps. One should not, therefore, think of the
half lumps as being scalar monopoles of opposite charge, generating a
dipole moment (recall that the half lumps are structures in the 
$\phi_3$ field, while the long-range field is dominated by $\phi_2$). 

Now consider the situation where two $B=1$ solitons with orientations
$\alpha_1,\alpha_2$ are placed at positions $\yv,\zv$ with $R=|\yv-\zv|$
large. The interaction energy that they experience should, on physical grounds,
coincide asymptotically with the interaction energy of the corresponding
scalar dipoles in the linear theory, that is,
\beq
E_{\rm int}=-\int_{\R^2}\kappa^{(1)}(\xv)\phi_2^{(2)}(\xv)d^2\xv
\eeq
where $\kappa^{(1)}(\xv)=\dv_1\cdot\nabla\delta(\xv-\yv)$ and $\phi^{(2)}$
is the field induced by dipole $\dv_2$ at $\zv$, that is
\beq
\phi_2^{(2)}(\xv)=-\frac{1}{2\pi}\dv_2\cdot\nabla\left(\log|\xv-\yv|\right).
\eeq
Hence
\bea
E_{\rm int}&=&\frac{1}{2\pi}\int_{\R^2}\dv_1\cdot\nabla[\delta(\xv-\yv)]
\dv_2\cdot\nabla(\log|\xv-\zv|)\, d^2\xv\nonumber\\
&=&\frac{1}{2\pi}\dv_1\cdot\nabla_y\, \dv_2\cdot\nabla_z\int_{\R^2}\delta(\xv-\yv)
\log|\xv-\zv|\, d^2\xv\nonumber \\
&=&
-\frac{1}{2\pi R^2}\left\{\dv_1\cdot\dv_2
-2(\nv\cdot\dv_1)(\nv\cdot\dv_2)\right\}
\eea
where $\nv=(\yv-\zv)/|\yv-\zv|$, the unit vector directed from dipole 2 to 
dipole 1.\, It is interesting to note that this is precisely
minus the interaction energy of a pair of electric dipoles in two-dimensional
electromagnetism, which can be understood by noting that, in scalar field 
theory, {\em like} monopole charges attract, while {\em opposite} charges
repel. In the case where the solitons are placed symmetrically about
the origin on the $x_1$ axis, that is, $\yv=(R/2,0)$, $\zv=(-R/2,0)$,
one sees that
\beq\label{asda}
E_{\rm int}=\frac{q_2^2}{2\pi R^2}\cos(\alpha_1+\alpha_2).
\eeq
This formula is rather remarkable. It has the interesting property that it
depends only on the {\em average} orientation $\alpha_1+\alpha_2$ of the
dipoles, and is independent of their phase difference $\alpha_1-\alpha_2$.
This is precisely the opposite of the situation found by Schroers et al
in the easy-axis model \cite{1995ZPhyC..65..165P},
\beq
E_{{\rm easy-axis}}=\frac{q_2^2}{2\pi}K_1(R)\cos(\alpha_1-\alpha_2),
\eeq
which highlights a fundamental difference between the two models. The
unit soliton in the easy-axis model is axially symmetric, and spatially
rotating the soliton is identical to internally rotating it (about the
axis $\phi_\infty$). Consequently, the interaction energy in the easy-axis
case must be invariant under $(\alpha_1,\alpha_2)\mapsto(\alpha_1+\Delta,
\alpha_2+\Delta)$, and hence can depend only on $\alpha_1-\alpha_2$. 
In the easy plane model, however, the unit soliton has no rotational symmetry.
One cannot rotate it internally about $\phi_\infty$ (the fields so obtained
do not solve the field equation), and spatially rotating the soliton produces
a genuinely physically distinct solution (e.g.\ its energy density changes).
So there is no reason to expect $E_{\rm int}$ to be independent of
$\alpha_1+\alpha_2$. That it depends {\em only} on $\alpha_1+\alpha_2$
is more surprising, and is a basic fact about dipole interactions in two
dimensions. In practice, it seems likely that higher-order corrections will
break the independence of $E_{\rm int}$ on $\alpha_1-\alpha_2$.

\begin{figure}
\begin{center}
\includegraphics[width=0.95\columnwidth]{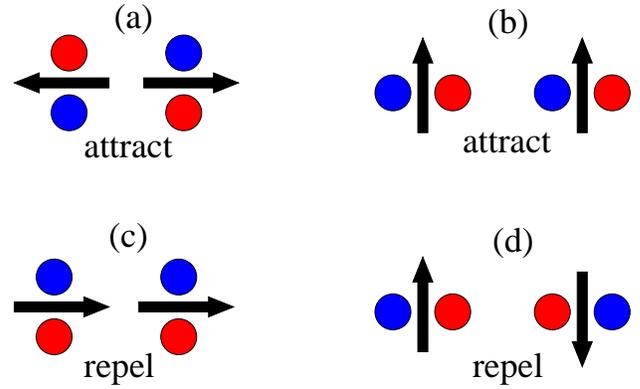}
\end{center}
\caption{Predicted behavior of two-Skyrmion superpositions. The disks
represent half lumps with $\phi_3>0$ (red) and $\phi_3<0$ (blue), while
the black arrows represent the scalar dipole moment associated with each
Skyrmion.}
\label{scjo}
\end{figure}

The most important consequence of formula (\ref{asda}) is that it predicts
long-range attraction between unit solitons if they are appropriately
oriented relative to one another, leading one to expect there to exist
stable, static, spatially localized solutions in all homotopy classes. In 
particular, soliton pairs initially oriented as in Fig.~\ref{scjo}(a) or
\ref{scjo}(b) are predicted to attract one another, while those
oriented as in Fig.~\ref{scjo}(c) or \ref{scjo}(d) are predicted to repel.
These predictions were vindicated by the numerical simulations 
described in the next section. 

\section{\label{sec:multiskyrmions}Multiskyrmions}

In this section, we report the numerical results on multi-Skyrmions, confirming the validity of the dipole
approximation, and give examples of the very complicated landscape of local minima of higher
multi-Skyrmions. We split this section in two according to two different ``branches'' in the shapes of the
minima. Before going further, it is worth noting several general observations.

First, the dipole approximation was confirmed for pairs of $B=1$ solitons. To confirm this, we used three
methods: the usual BFGS method, simple gradient flow and also a Newton flow second order dynamics, where we
set $\psi := \partial_0 \phi$ and $\partial_0 \psi = -\partial_\phi E, \partial_0 \phi = \psi$. They all agree
that two unit Skyrmions starting in the attractive channel coalesce into a $B=2$ multisoliton with an energy
lower than twice that of the unit Skyrmion. The resulting two-Skyrmion is depicted in
Fig.~\ref{fig:minimizers}\subref{fig:b2}. However, confirming the repulsive channel is slightly more
challenging. First, the computational box needs to be large enough, otherwise the boundaries will exert
pressure on the solitons and they react by rotating into the attractive channel and coalescing. Second, the
solitons need to be placed exactly symmetrically on the lattice, lest they again rotate into the attractive
channel and coalesce. When set up correctly, however, placing two $B=1$ Skyrmions in the repulsive channel
results in them receding. Thus we can confirm the qualitative correctness of the dipole approximation.

Second, all the results presented below are done in a standard lattice with $h=0.1$ and $N=801$, except the
case $B=32$, which does not fit into the smaller lattice at all. For those values of $B$, where multiple
minima were found, two additional methods were used for their initial configurations. The first method is a
simple cut and paste of other solitons, for example, two $2$-solitons were put side by side to produce a
$4$-soliton. The other method was to start with $N=801, h=0.05$ which resulted in a different minimum than
$N=801, h=0.1$. This is due to boundary pressure: if $N$ was doubled, the other result reemerged. When this
happened, the $h=0.05$ solution was made sparser by a factor of 4 to put it into the standard lattice, where
it was then minimized again to get a result which can be compared with the others. Where several initial
states produced the same minimizer, we only mention one of the initial states.

\subsection{\label{subsec:b_2_6}Charges $B \in [2,6]$}

Minimal energy Skyrmions of charges $B$ in the interval $[2,6]$ found using our numerical scheme form regular
polygons, where there are half lumps at the $2B$ vertices. The energies of these minimizers are always lower
than $BE_1$, where $E_1$ is the energy of the unit Skyrmion, so there is good reason to believe they are
stable in accordance with the above dipole picture. It should be noted, however, that the symmetry of these
minimizers is the same as that of the initial data $w=\lambda (x + iy)^B$. Therefore, one is led to suspect
that the random ripples added to $w$ may not be enough to break the symmetry. It is therefore necessary to
introduce a stronger symmetry break. For $B=2$, the computations done to confirm the dipole approximation
provide this break, and still relax to the same square final configuration. The obvious way to test the higher
charges was therefore to place two lower charge solitons side by side and minimize their energy.

Indeed, for charges 5 and 6, there are minima, which were only found using this spliced initial
configuration. However, they have higher energy than the respective polygons, and therefore represent a local
minimum only. For the charges $3$ and $4$, various different initial configurations were tried, but they
always relaxed to the respective polygon. We present the minimizer of $B=6$ as an example of the polygon
shaped minimizers in Fig.~\ref{fig:minimizers}\subref{fig:b6}.

\subsection{\label{subsec:b_8_}Charges $B > 6$}

The $2B$-gon form of the global minimizers only persists up to $B=6$ after which, the shape of the minimizer
is completely different and there is no obvious rule for the shape. We will briefly discuss these minimizers
below, but first it should be noted that the symmetric (plus random ripples) initial state still produces a
polygon shape local minimizer for all charges except 13 and 14, but this is no longer a global minimum. The
polygon exists for charges 13 and 14, also, but was only obtained by starting the minimization process with
$N=399,h=0.1$ and embedding the result in the standard lattice. The polygons are, however, quite stable and
the energy differences between them and the global minima are very small. For example, we squeezed some of the
polygons to half the size in the $y$ direction and they relaxed back to the polygon shape instead of the
global minimum.

The solutions obtained from lower charge solitons set side by side and starting the minimization in a
$N=801,h=0.05$ lattice, produce a veritable zoo of different minimizers, of which all but one have lower
energies than the $2B$-gons, the exception being charge 14. Because of the amount of work involved in patching
the solitons together and the total number of computations already performed, we did not attempt an exhaustive
search with side-by-side-type initial configurations of all charges.

\begin{figure}[h]
  \centering
  \subfloat[][Global minimizer for $B=2$.]{\includegraphics[width=0.4\columnwidth]{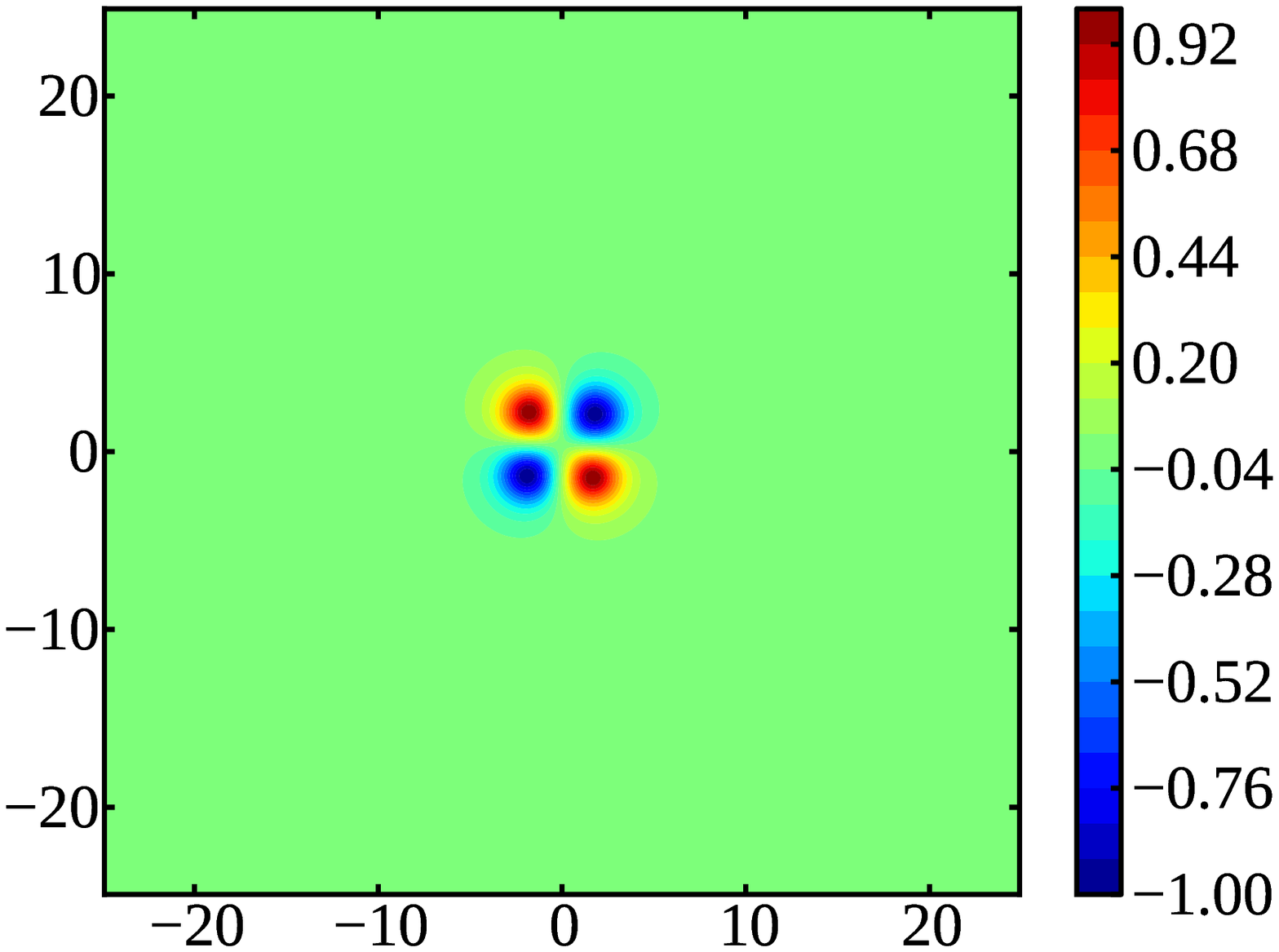}\label{fig:b2}}
  \subfloat[][Global minimizer for $B=6$.]{\includegraphics[width=0.4\columnwidth]{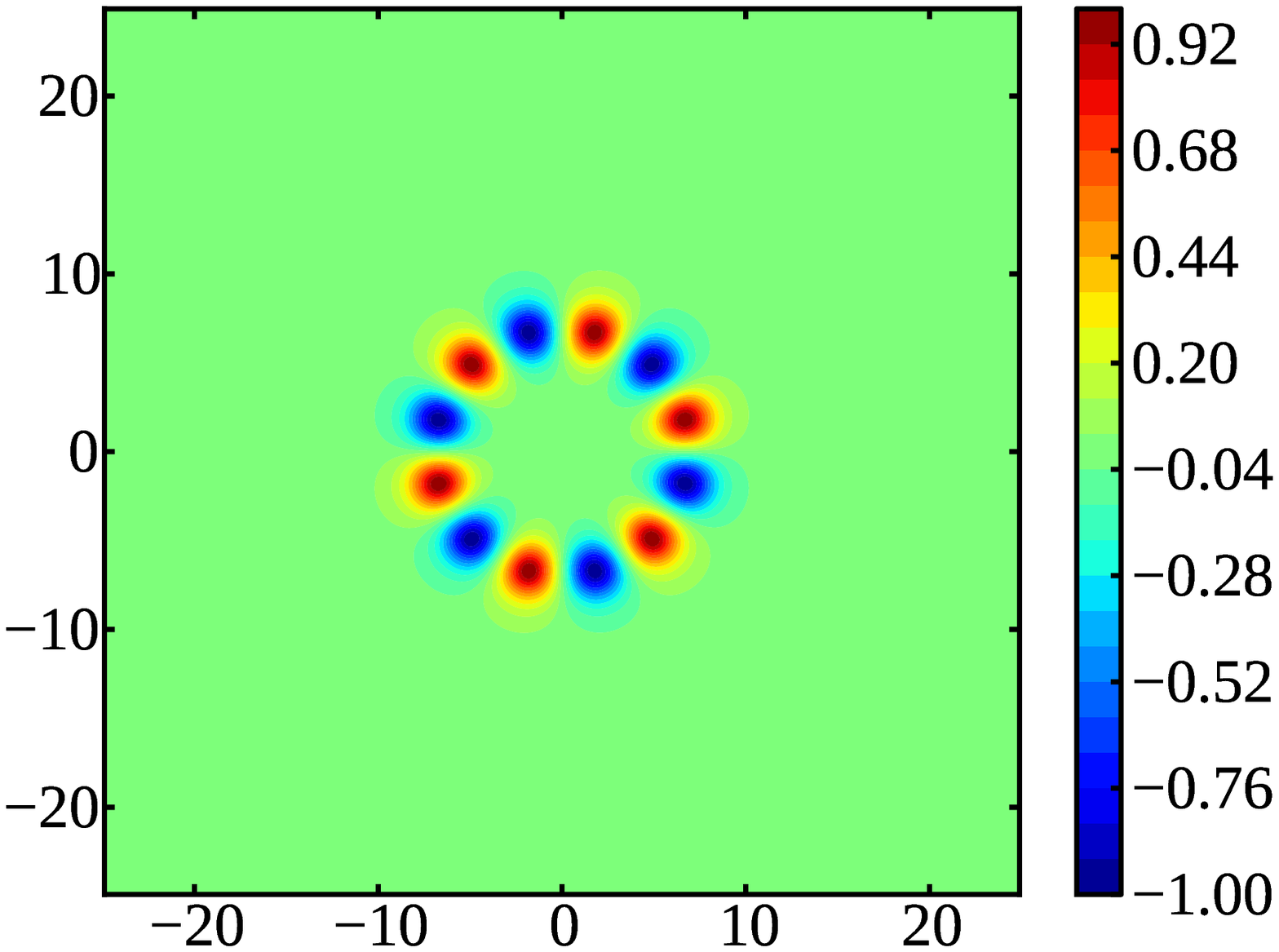}\label{fig:b6}}\\
  \subfloat[][Global minimizer for $B=7$.]{\includegraphics[width=0.4\columnwidth]{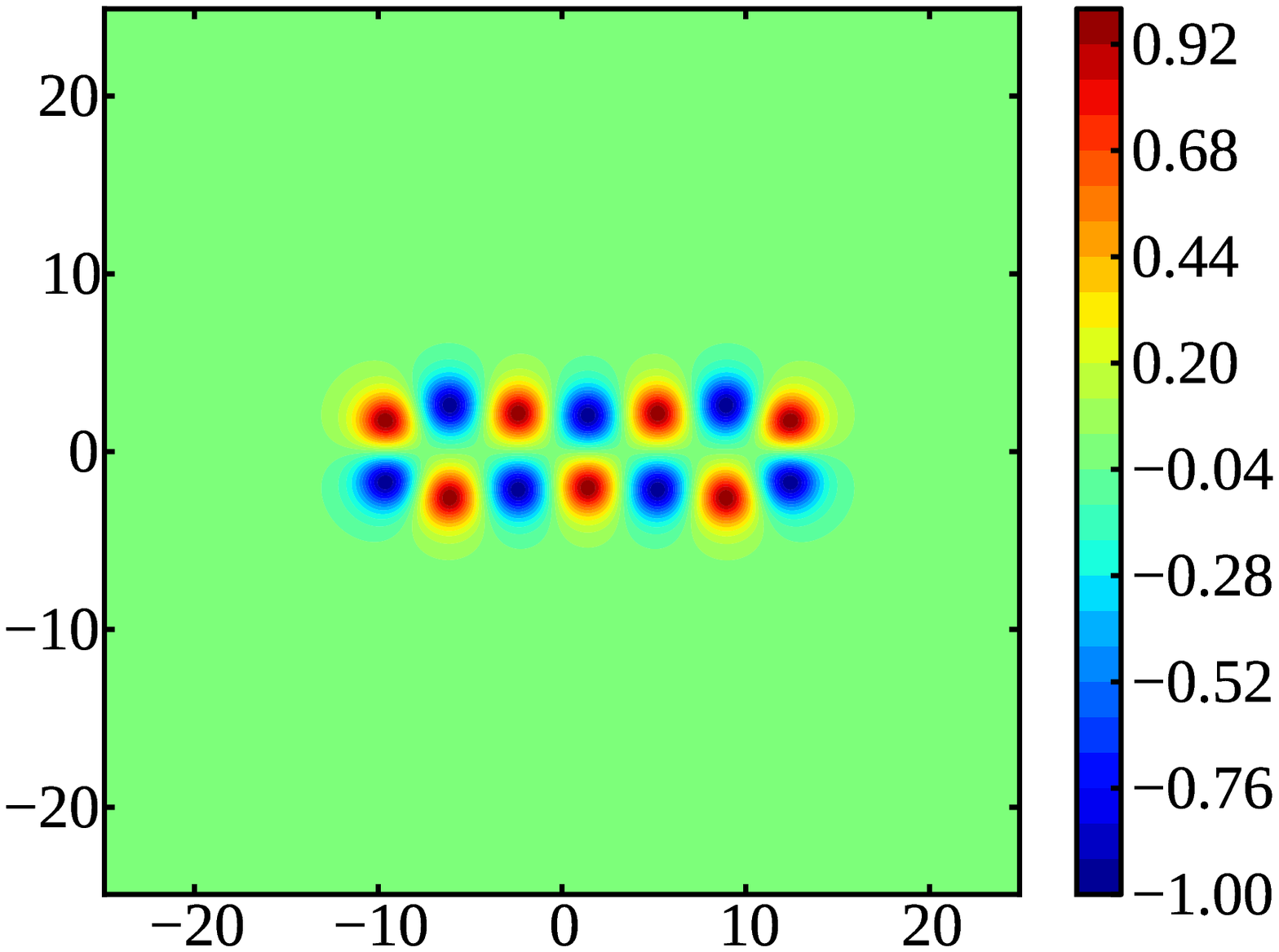}\label{fig:b7}}
  \subfloat[][Global minimizer for $B=8$.]{\includegraphics[width=0.4\columnwidth]{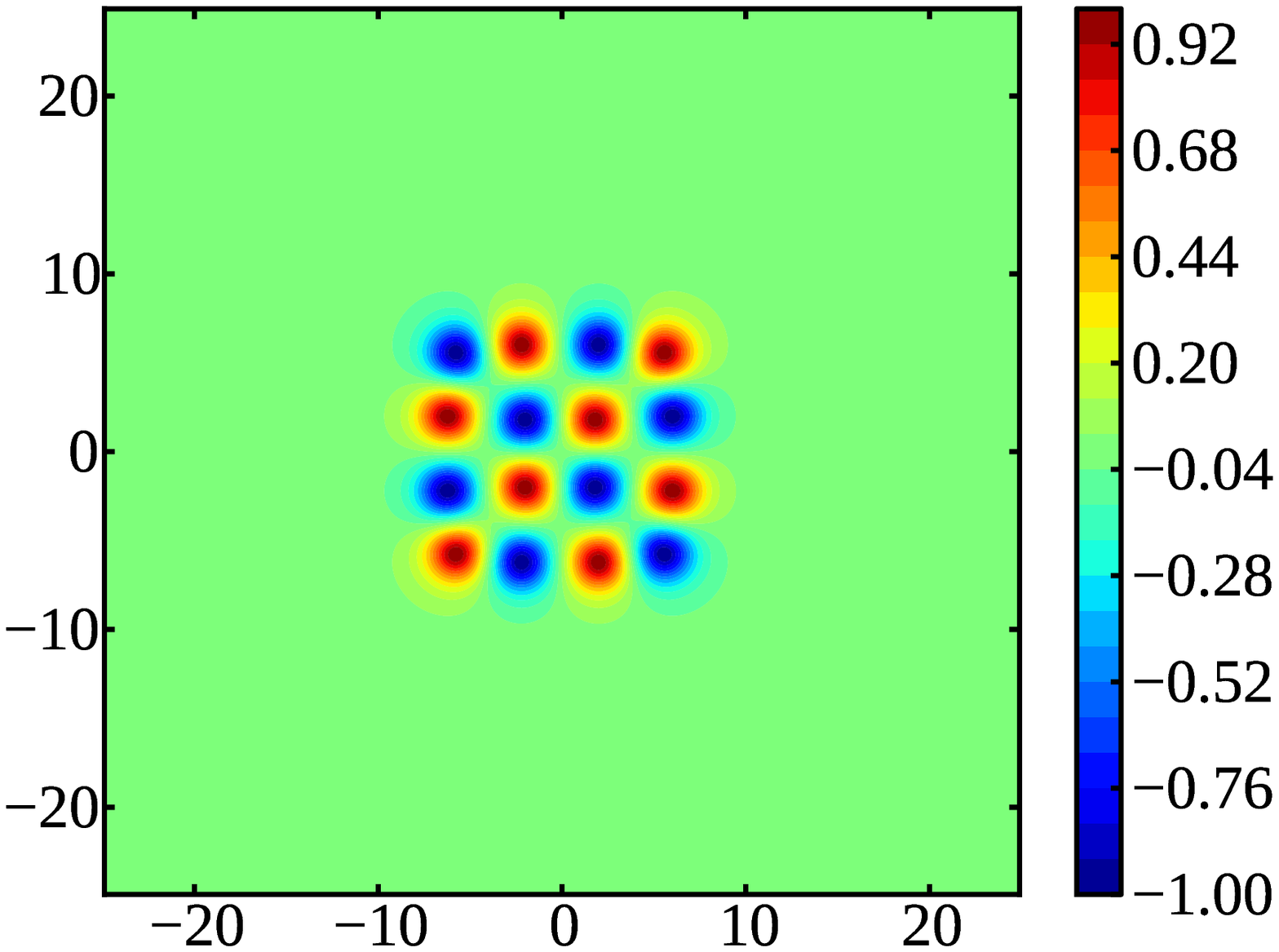}\label{fig:b8}}\\
  \subfloat[][Global minimizer for $B=18$.]{\includegraphics[width=0.4\columnwidth]{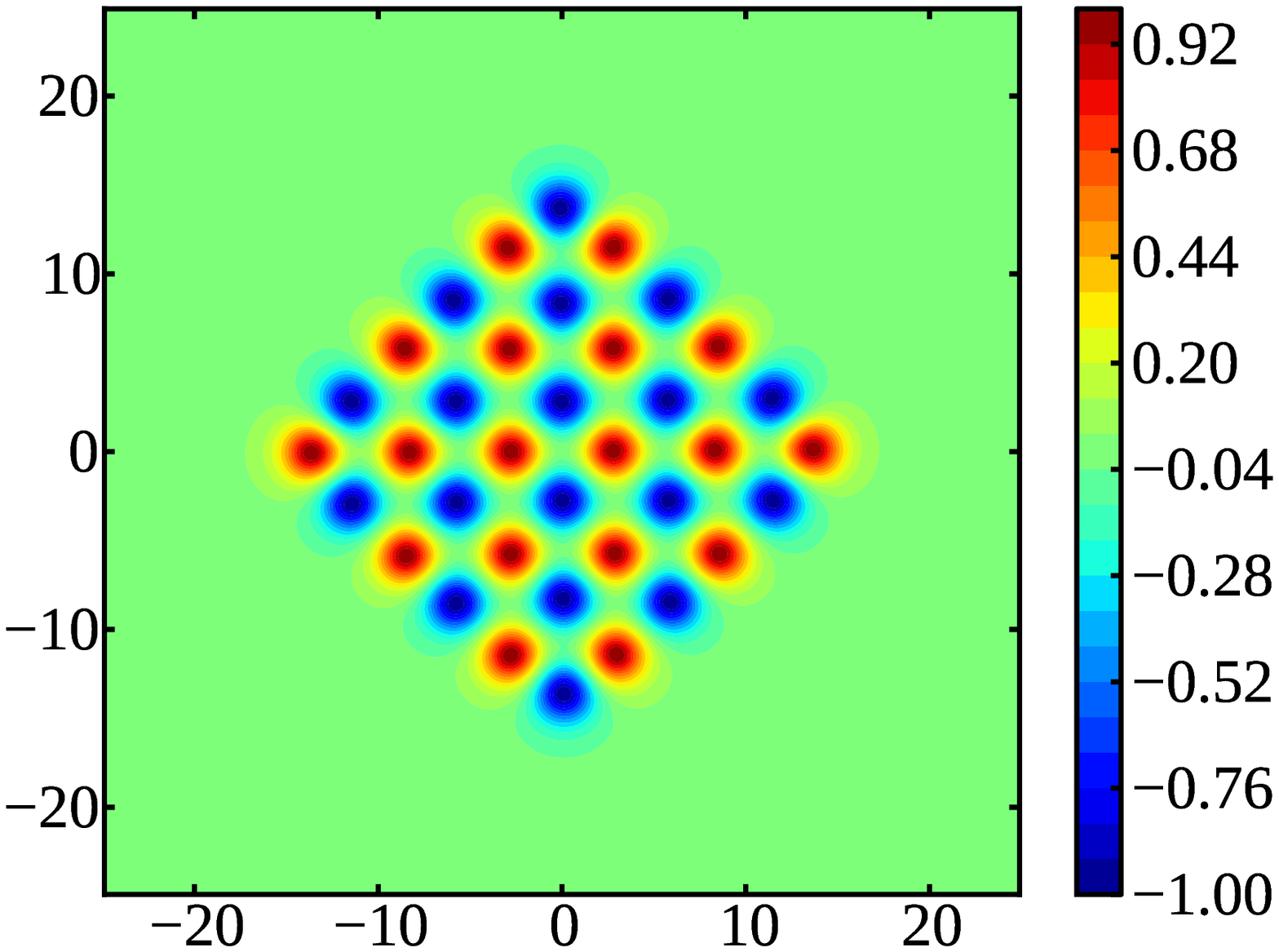}\label{fig:b18}}
  \subfloat[][Global minimizer for $B=32$.]{\includegraphics[width=0.4\columnwidth]{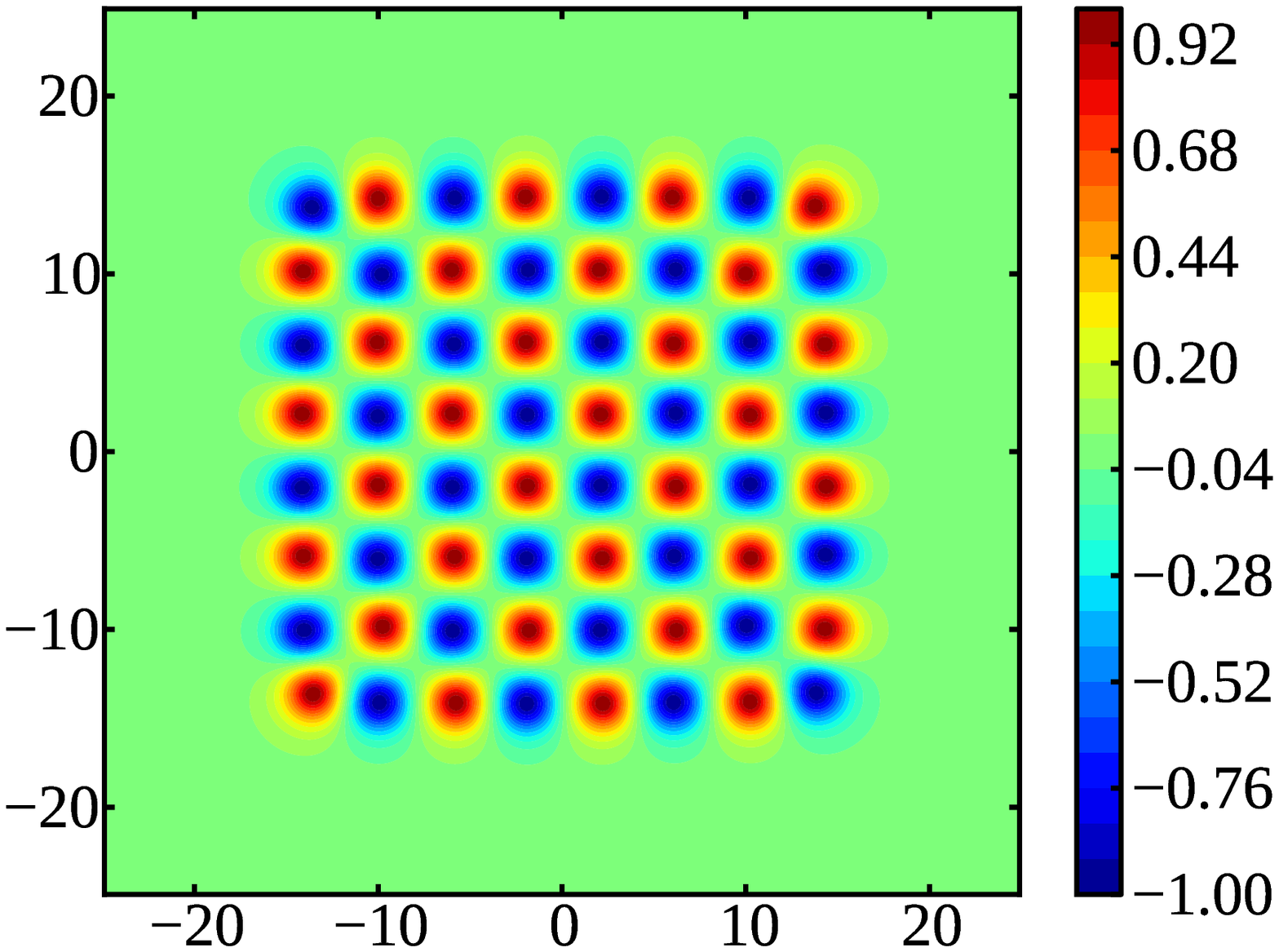}\label{fig:b32}}\\
  \subfloat[][Global minimizer for $B=13$.]{\includegraphics[width=0.4\columnwidth]{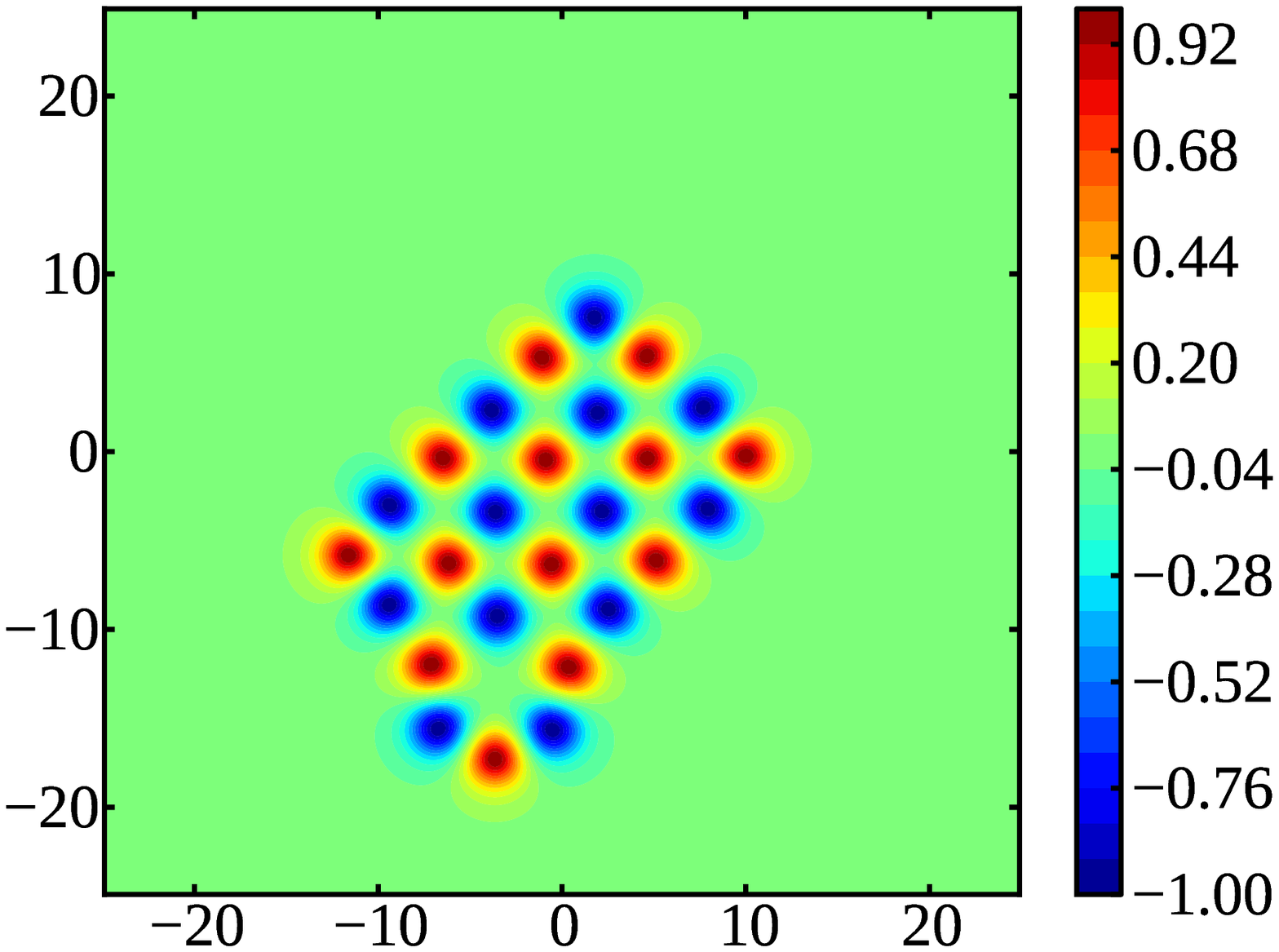}\label{fig:b13}}
  \subfloat[][Local minimizer for $B=14$.]{\includegraphics[width=0.4\columnwidth]{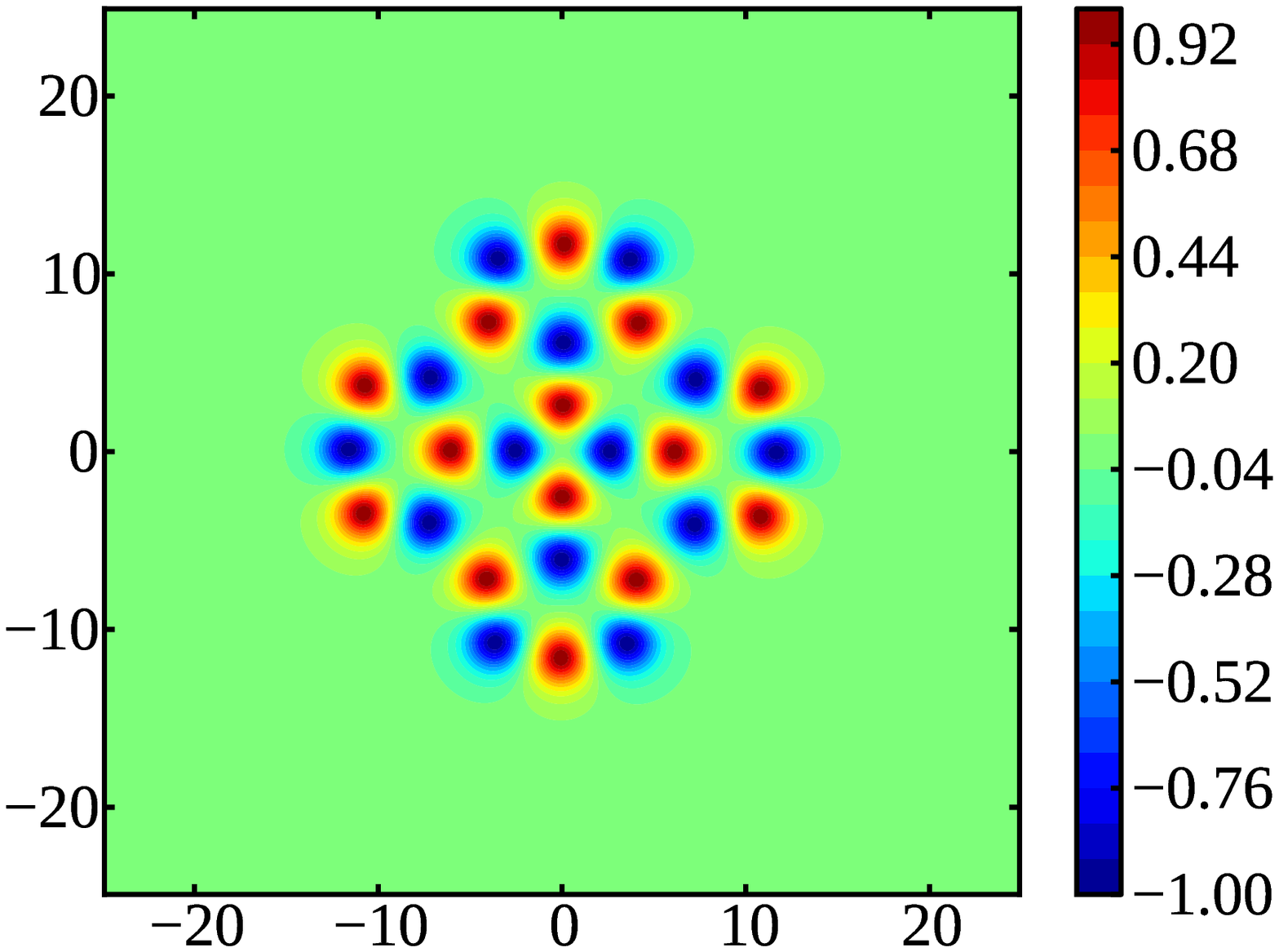}\label{fig:b14}}
  \caption{Sample minimizers for higher degrees. All plots are contour plots of $\phi_3$.}\label{fig:minimizers}
\end{figure}

The first nonpolygon minimizer occurs at $B=7$ and looks like a slightly distorted rectangle of $2 \times 7$
half-lumps. It is depicted in \ref{fig:minimizers}\subref{fig:b7} and was obtained by putting a $B=3$ soliton
next to a $B=4$ one. A similar minimizer is also found for $B=9$ using a $B=4$ soliton beside a $B=5$ soliton
as the initial configuration. These are global minima, within the limits of our numerics.

At $B=8$ the half lumps of the minimum energy configuration arrange themselves into a square lattice of $4
\times 4$ half lumps, as seen in Fig.~\ref{fig:minimizers}\subref{fig:b8}. The initial configuration consisted
of four $B=2$ solitons at the vertices of a square. Similar square lattice configurations are also found for
$B=18$ and $B=32$, shown in Figs.~\ref{fig:minimizers}\subref{fig:b18} and
\ref{fig:minimizers}\subref{fig:b32}. The initial configurations were as follows. For $B=18$, two $B=4$ and
two $B=5$ solitons were put at the opposite vertices of a square and for $B=32$, four $B=8$ solitons were
placed at the vertices of a square. It is interesting to note that the $B=2$ solution can also be thought of
as $2 \times 2$ square lattice of half lumps, providing us with a tantalizing opportunity to conjecture that
all $2B=(2q)^2, q\in\integers$ will fall into similar square lattices. While these are all possible square
lattices for $B \le 32$, this raises the question of whether there are other symmetric minimizers and what are
their symmetries, if any.

It seems that using various spliced initial configurations, the square symmetry can be relaxed to produce
other rectangular shapes. We have already seen two of these at $B=5$ and $B=6$, which are not global minima
and at $B=7$ and $B=9$, which are the global minima for these charges. For all four cases, the short edge was
the shortest possible, but for $B=10,\; 12, \text{ and } 14$ a more squarelike rectangular lattice of $4
\times 5$, $4 \times 6$, and $4 \times 7$ half lumps, respectively, exists. These are all global minima; they
were all obtained using spliced initial configurations.

The global minima for $B=11$ and $B=13$ have the shape of a distorted rectangle, where the distortion consists
of an extra two half lumps in one corner. These were both achieved by using $N=801,h=0.05$ initial states (as
always, these were later coarsened and embedded into the standard lattice). Of these, we show the minimizer
$B=13$ in Fig.~\ref{fig:minimizers}\subref{fig:b13} as an example.

As the charge increases, finding the global minimum -- or what can reasonably be believed to be one -- becomes
increasingly difficult. It is likely that the (possibly distorted) rectangular shapes prevail as global minima
for charges above $B=14$, but we have not been able to ascertain this.
As an example of how difficult it is to find the global minimum, consider charge $B=9$. There is a local
minimum of the distorted rectangle type with two extra half lumps in one corner. To converge into this
configuration, an initial configuration of charges $B=1$ and $B=8$ is needed, but it is not enough to just put
these side by side. They must be placed at very precise locations in the lattice. Although this arrangement is
natural, place the $B=1$ solution outside the $B-8$ solution along one of its diagonals, it is very difficult
to guess the correct arrangement for higher charges. Moreover, at higher charges, there is an ever increasing
number of different combinations to choose from.

In addition to the global and polygon-shaped minima, we found some other local minima. Most of these have
energies between the global minimum and the polygon and have various (or no obvious) symmetries, like the
distorted rectangle of $B=9$ mentioned above. The one notable exception is the lucky cloverleaf at $B=14$,
whose energy exceeds that of polygon and has an obvious symmetry. It is depicted in
Fig.~\ref{fig:minimizers}\subref{fig:b14}.

We now briefly return to the case $B=6$. Since it is the last one where the polygon seems to be the global
minimum, it deserves special attention: how can we be certain some other configuration does not give a lower
energy? Since it seems conceivable that a rectangular lattice of $3 \times 4$ half lumps could exist, we tried
several methods of producing it: we split the rectangular $B=12$ minimizer in two, keeping just one half;
spliced together all possible combinations of two lower charge solutions side by side and even spliced
together two $B=1$ and two $B=2$ solitons in a crosslike configuration. They all relax either to the polygon
or the $2 \times 6$ rectangle. Of course, it is possible, that the rectangles reported above for the higher
charge solitons might not be the true global minima, but we believe they are: any other rectangle would
necessarily have a bigger difference between its long and short sides. Therefore, we did not attempt to find
any other rectangles.

\begin{figure}[h]
  \centering
  \includegraphics[width=\columnwidth]{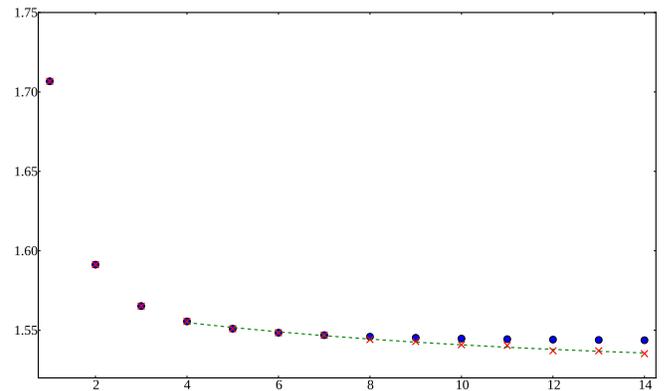}
  \caption{Normalized energy per $4 \pi B$ versus $B$ of those configurations which we believe are global
    minima (crosses), the polygon-shaped solitons (circles) and a least squares fit of the form $a+b\exp{c
      B}$ (dashed curve).}
  \label{fig:EperB}
\end{figure}

Finally, we summarize the charges, normalized energies per $B$ and shapes of our global minimizers in
Table~\ref{tab:b1-32}. We also conjecture an exponential decay law for $E(B)/(4 \pi B)$ of the global
minimizer which is approaching the value $1.52$ when $B$ increases without limit. This behavior, along with a
least squares fit of the form $a+b\exp{c B}$, is displayed in Fig.\ref{fig:EperB}.

\begin{table}[h]
  \centering
  \begin{tabular}{|l|c|c|c|}
    \hline
    B & Global minimum E/B & Polygon E/B & Shape \\
    \hline
      1 & 1.707 & 1.707 & pair of half lumps \\
    \hline
      2 & 1.591 & 1.591 & square \\
    \hline
      3 & 1.565 & 1.565 & regular polygon \\
    \hline
      4 & 1.556 & 1.556 & regular polygon \\
    \hline
      5 & 1.551 & 1.551 & regular polygon \\
    \hline
      6 & 1.548 & 1.548 & regular polygon \\
    \hline
      7 & 1.547 & 1.547 & rectangle \\
    \hline
      8 & 1.544 & 1.546 & square \\
    \hline
      9 & 1.543 & 1.545 & rectangle \\
    \hline
     10 & 1.541 & 1.545 & rectangle \\
    \hline
     11 & 1.540 & 1.544 & distorted rectangle \\
    \hline
     12 & 1.537 & 1.544 & rectangle \\
    \hline
     13 & 1.537 & 1.544 & distorted rectangle \\
    \hline
     14 & 1.535 & 1.544 & rectangle \\
    \hline
    \hline
    18 & 1.532 & 1.543 & square \\
    \hline
    \hline
    32 & 1.526 & 1.543 & square \\
    \hline
  \end{tabular}
  \caption{Energies of the baby Skyrmions}
  \label{tab:b1-32}
\end{table}

\section{\label{sec:concluding_remarks}Concluding remarks}

We have used numerical minimization to show that multi-Skyrmions in the easy plane baby Skyrme model exist and
are stable. For each charge above 4, we have found at least two minima, sometimes three or even four. Of
these, we have most likely identified the global one for most charges in the range $[1,14]$, $18$, and
$32$. These minima exhibit various different types of symmetries and it is conceivable that a method exists
for determining the shape of the minima without finding the minima themselves. However, this remains an open
question. The prevalence of rectangles or slightly distorted rectangles in this work might provide a way to
answer the question. We also present a conjecture that $\lim_{B\to\infty}E(B)/(4\pi B) \approx 1.52$.

Like baby Skyrme models with some other potentials, we have identified an attractive and repulsive channel for
two well-separated unit Skyrmions using a dipole approximation and confirmed its validity by numerical
simulations.

The rectangular lattices formed by the half lumps are reminiscent of condensed matter systems, where (baby)
Skyrmions have been experimentally observed (see, for example, \cite{2009Sci...323..915M,
  2010Natur.465..901Y}). These are not, however, modeled by the baby Skyrme model, so it would be interesting
to investigate whether the easy plane baby Skyrme model could be used as an effective model for some condensed
matter systems.

\begin{acknowledgments}
  This work was supported by the UK Engineering and Physical Sciences Research Council. JJ also wants to thank
  Lisandro Dalcin for help with the \emph{tao4py} library.
\end{acknowledgments}

\bibliography{bibliography}

\end{document}